# Positioning of a Next Generation Mobile Cell to Maximise Aggregate Network Capacity

Paulo Furtado Correia[1][0009-0006-9504-5886], André Coelho[2][0000-0003-4910-4567] and Manuel Ricardo[3][0000-0003-1969-958X]

[1] paulo.j.correia@inesctec.pt
[2] andre.f.coelho@inesctec.pt
[3] manuel.ricardo@inesctec.pt
INESC TEC and Faculdade de Engenharia, Universidade do Porto, Portugal

**Abstract.** In wireless communications, the need to cover operation areas, such as seaports, is at the forefront of discussion, especially regarding network capacity provisioning. Radio network planning typically involves determining the number of fixed cells, considering link budgets and deploying them geometrically centered across targeted areas. This paper proposes a solution to determine the optimal position for a mobile cell, considering 3GPP pathloss models. The obtained position for the mobile cell maximises the aggregate network capacity offered to a set of User Equipments (UEs), with gains up to 187% compared to the positioning of the mobile cell at the UEs' geometrical center. The proposed solution can be used by network planners and integrated into network optimisation tools. This has the potential to reduce costs associated with the radio access network planning by enhancing flexibility for on-demand deployments.

## 1 Introduction

Industrial environments should provide seamless wireless connectivity throughout operational areas, enabling communications among various devices, diverse machinery, and personnel. This paper studies the added value of a Next Generation Mobile Cell (NGMC) in a seaport container yard, where ground and sea operations are bustling with variable demand for wireless connectivity. Facilities are typically used for storage, handling and maintenance of shipping containers, where ships, mobile cranes and trucks operate. NGMCs have the unique ability to move within seaport areas, carried by automated guided vehicles or Unmanned Aerial Vehicles (UAVs), offering enhanced capacity and improved coverage for UEs. The complex and dynamic environment and the variable traffic demand within the port, including tractor boats and mobile cranes or trucks taking and stacking containers, require on-demand wireless connectivity. In order to address these challenges, tailored network planning and further deployment approaches are crucial to provide communications with adequate Quality of Service. Optimising NGMC positioning includes multiple factors to consider, including container stacking areas, truck movement patterns, and key operational zones. Radio coverage should ensure network capacity maximisation to improve communications performance and enhance the overall productivity of port operations' services within smart ports [1]. Moreover, legacy networks can coexist with NGMCs operating in different frequency bands, such as Frequency Range 1 (FR1) and Frequency Range 2. This paper focuses on FR1 and provides the following original contributions:

1. An optimisation problem formulation to obtain the 3D positions for NGMCs, maximising the aggregate network capacity over an outdoor area with multiple UEs. 3GPP recommends an error model to compute the block error rate and manage hybrid automatic repeat request (HARQ) retransmissions. It does this based on Signal-to-Interference-plus-Noise Ratios (SINRs) per resource block (RB) historical data. Developed according to standard link-to-system mapping techniques, this model is an exponential effective signal-to-noise ratio mapping (EESM), originally created for single-antenna transmissions. We have simplified this 3GPP error model using a heuristic that provides a spectral efficiency regression line as a function of SINR, while maintaining the link-level performance accuracy;
2. A network performance evaluation study conducted using MATLAB and the ns-3 network simulator, employing the proposed optimisation problem and considering randomly generated scenarios. The performance evaluation study considers a specific seaport configuration, as a representative example. However, the contributions are applicable to any other outdoor scenario.

The remainder of this paper is structured as follows. Section 2 provides a literature review, presenting the related work on the positioning of NGMCs across outdoor areas such as seaports. Section 3 introduces the system model, including a description of the main constraints, methodology, and performance metrics employed. Section 4 presents the performance evaluation for diverse scenarios, including an analysis of the results obtained. Finally, Section 5 summarises the key findings and identifies directions for future work.

## 2 Related Work

NGMCs have been proposed in the literature since 2015, enabling multiple Heterogeneous Networks to improve network throughput, extend wireless coverage, and offload network traffic in multiple use cases [14]. Typical radio and network planning optimisation best practices, conducted by Mobile Network Operators (MNOs) or subcontracted entities, are based on link budgets and the number of fixed cells to deploy, geometrically referenced to the center of each coverage area. This procedure is described in [20]-[21], focusing on optimising data rates by leveraging Carrier Aggregation with radio planning. However, optimal cell positioning is not addressed. [12] explores the optimal UAV 3D position carrying a 28 GHz cell. The study was improved to a 3D case based upon [13], incorporating actual, fine-grained terrain data and accounting for location-dependent line-of-sight (LoS) conditions measured on the fly, resulting in higher gains by avoiding non-line-of-sight (NLoS) conditions and considering UAV heights. However, it does not present simulation results and does not consider 3GPP-compliant pathloss models. Integrating visual information and obstacle avoidance mechanisms, alongside mathematical analyses of capacity bounds, have enriched connectivity solutions [2]-[4]. [2] proposes an obstacle-aware wireless connectivity for 5G UEs, leveraging a novel on-demand mobility management function, which enables the remote positioning of a communications mobile robotic platform by using visual information from on-board video cameras. [3] studies network capacity bounds expressed as the maximum number of users. It considers cell interference and limited uplink transmission power, while assessing the performance of 5G links in different propagation environments using the extended COST-231 Hata model. [3] and [4] leverage context information related to user positions, using a combined Control/User-plane split to improve the directional cell discovery process through information gathered over time. Moreover, considerations for different backhaul link types and user-centric approaches, rather than network-centric, have been explored to optimise served UEs and backhaul sum rates. However, no placement accuracy and capacities are detailed [5]. In Integrated Access and Backhaul (IAB) deployments, studies have optimised precoder designs, UE-Base Station (BS) associations, and UAV hovering locations to maximise network sum rates or spectral efficiencies [6]-[8]. In city centers, IAB solutions positioned in buses or trains can improve passengers' connections in terms of throughput and latency on uplink and downlink [11]. However, IAB still lacks further standardisation and existing studies do not consider network capacities. [9] focuses on ultra-reliable low latency communications users, optimising drone base station locations, user associations, and backhaul bandwidth allocations for delay-sensitive applications while maximising sum rates. Wireless backhaul optimisation between fixed small cells and UAV cells, adaptive modulation and coding (AMC) techniques, and traffic-aware UAV placement algorithms have been proposed to maximise system throughput and facilitate controlled topology and backhaul communications paths [17]-[19]. [17] proposes the optimal placement for a UAV, in order to maximise the sum rates of the whole network, consisting of UAV cells and terrestrial small cells. However, the work focuses on wireless backhaul. In contrast, [18] formulates the problem of maximising the number of covered UEs, subject to bit rate requirements, for a certain capacity limit while mentioning the absence of capacity maximisation studies in the literature. [19] focuses on uplink throughput maximisation using a reinforcement learning multi-agent proportional fair scheduling in the reward function. In [10], communications strategies for Sidelink are studied to convert NLoS to LoS conditions to improve outage probability in smart factory indoor environments; however, this can also apply to smart ports. Sidelink refers to a communications mode where devices in close proximity to each other establish a direct wireless connection. Flying networks, consisting of access points or cellular BSs, can also leverage traffic-aware UAV placement algorithms to enable controlled topology and backhaul communications paths with sufficient capacity. Within this context, [15] and [16] address the placement of a single UAV using Wi-Fi.

Overall, through multiple works spanning NGMCs, urban simulations, UAV deployments, optimisation techniques, and smart factory environments, the literature addresses network access and backhaul sum rates optimisation with micro or macro cell positioning, including network performance evaluations. Additionally, combined factors, such as resource allocation, interference, transmission power and obstacles' interposition are also shown how they can impact results. However, the existing works do not focus on aggregate capacity, as addressed in this paper, but mostly on sum rates.

## 3 System Model

Let us consider a three-dimensional networking scenario where NGMCs must be placed to provide wireless connectivity to multiple UEs. The problem involves positioning these NGMCs and determining their associations with UEs. In order to obtain the Euclidean coordinates for placing them ($x_m^M, y_m^M, z_m^M$) and respective user associations ($S_{j,m}^M$) offered to the UEs, an optimisation problem is formulated herein.

### 3.1 Problem Formulation

The problem is formulated as a Mixed-Integer Nonlinear Programming problem, with the objective function given by (1). It maximises the sum of the network capacities offered to the UEs with single user associations, ensured by (2), and whose 2D distances (3) and respective applicable spectral efficiencies (4) align with [22]-[24]. Although this paper considers only a single NGMC, the problem formulation is generic and can be applied to $M$ NGMCs. Table 1 lists the main notations used in the problem formulation.

$$\underset{x_m^M, y_m^M, z_m^M, S_{u,m}^M}{\text{maximise}} \quad \sum_{u \in U} \sum_{m \in M} (C_{u,m}^M S_{u,m}^M) \tag{1}$$

Subject to:

$$\sum_m S_{u,m}^M = 1, \forall u \in [1, U] \tag{2}$$

$$10 \leq d_{2Du,m}^M \leq 5000, \forall u \in [1, U], \forall m \in [1, M] \tag{3}$$

$$0 \leq SE_{u,m}^M \leq 6.4, \forall u \in [1, U], \forall m \in [1, M] \tag{4}$$

**Table 1.** List of the main notations used in the problem formulation.

| Notation | Definition |
| --- | --- |
| $u, m$ | Symbols representing UE and NGMC in volume $V$. |
| $NGMC_m$ | NGMC $m$, out of $M$. |
| $UE_u$ | User equipment $u$, out of $U$. |
| $V$ | Volume of coverage defined by 3D Cartesian coordinates between minimum (-1000 m) and maximum (1000 m) values. |
| $C_{u,m}^M$ | Capacity provided by $NGMC_m$ to the $UE_u$, in bits per second (bit/s). |
| $B$ | Number of RBs used per numerology. |
| $\Delta f_{r1}$ | Subcarrier spacing (SCS) for FR1, in Hz. |
| $U_m$ | Number of UEs associated to $NGMC_m$. |
| $S_{u,m}^M$ | Binary variables indicating user association of $UE_u$ to $NGMC_m$, where '1' means associated and '0' otherwise. |
| $Bw_{u,m}^M$ | Bandwidth of the channel between $UE_u$ and $NGMC_m$, in Hz. |
| $SINR_{u,m}^M$ | SINR of the channel between $UE_u$ and $NGMC_m$, in dB. |
| $R_{u,m}^M$ | Reference Signal Received Power (RSRP) of $UE_u$ with respect to $NGMC_m$, in dB. |
| $N_{fr1}$ | Thermal noise power for FR1, in dB. |
| $G_t$, $G_r$ | Transmitter and receiver antenna gains, in dBi. |
| $P_t$ | Transmission power, in dBm. |
| $c$ | Speed of light in vacuum ($3 \times 10^8$ m/s). |
| $f_m^M$ | Carrier frequency, in GHz. |
| $x_m^M, y_m^M, z_m^M$ | Euclidean coordinates of $NGMC_m$, in m. |
| $x_u^U, y_u^U, z_u^U$ | Euclidean coordinates of $UE_u$, in m. |
| $h_{BS}, h_{UT}$ | Heights of $NGMC_m$ and $UE_u$, in meters. $h_{BS}$=25 m; $h_{UT}$=1.5 m, as defined in Table 7.4.1-1 of [23] for 3GPP Urban Macro (UMa) scenario. |
| $d_{2Du,m}^M, d_{3Du,m}^M$ | Euclidean 2D and 3D distances between $UE_u$ and $NGMC_m$, in meters. |
| $d_{BPu,m}^M$ | Euclidean 2D Breakpoint distances between $UE_u$ and $NGMC_m$, in meters. |
| $\mu_{fr1}$ | Numerology used for FR1. |
| $\eta_{u,m}^M$ | Spectral efficiency of the established downlink Radio Network Temporary Identifier (RNTI) between $UE_u$ and $NGMC_m$ in bit/s/Hz. |
| $SE_{u,m}^M$ | Obtained regression for spectral efficiency of the established downlink RNTI between $UE_u$ and $NGMC_m$, as a linear function of the $SINR_{u,m}^M$, in bit/s/Hz. |
| $PL_{UMa\_LOSu,m}^M$ | Pathloss for UMa scenarios considered for wireless link between $UE_u$ and $NGMC_m$. |

The capacity of the RNTI between $UE_u$ and $NGMC_m$ is introduced in (5).

$$C_{u,m}^M = \frac{B \ \Delta fr1 \ \eta_{u,m}^M}{U_m} \tag{5}$$

It depends on the spectral efficiency defined in (6), which is always positive. It also considers the SCS of the FR1 frequency band, as defined in (7), the number of RBs, and the number of user associations to the same NGMC, as presented in (8). The bandwidth used per RNTI, representing the direct wireless link between $UE_u$ and $NGMC_m$, is given by (9).

$$\eta_{u,m}^M = \begin{cases} SE_{u,m}^M, & SE_{u,m}^M \geq 0 \\ 0, & SE_{u,m}^M < 0 \end{cases} \tag{6}$$

$$\Delta fr_1 = 2^{\mu_{fr1}} .15 \text{ kHz} \tag{7}$$

$$U_m = \sum_{u \in U} S_{u,m}^M \tag{8}$$

$$Bw_{u,m}^M = \frac{B \ \Delta fr1}{U_m} \tag{9}$$

Equation (10) represents the spectral efficiency as a linear regression obtained as a function of $SINR_{u,m}^M$ (dB), as shown in Fig. 1.

$$SE_{u,m}^M = 0.23 \ SINR_{u,m}^M - 0.21 \tag{10}$$

The SINR intervals, ranging from minimum to maximum values, for each Modulation and Coding Scheme (MCS), form the basis of the EESM error model used in next generation radio communications. According to the MCS index table 2 in [22] for PDSCH, the MCS index is related with modulation order, target code rate and respective spectral efficiency. [24] also discusses these SINR intervals for both NR PHY LDPC base graph types 1 and 2, for different code block sizes.

The spectral efficiency chart, presented in Fig. 1, which is a function of SINR in dB, consists of a set of values and a linear regression. These values represent the middle points of the SINR intervals for each MCS, each associated with a corresponding spectral efficiency. The linear regression in (10) is also depicted in Fig. 1. It was obtained to simplify the abstraction model and approximate the spectral efficiency using a least squares method. When (10) is applied in (6), it provides the spectral efficiency $\eta_{u,m}^M$. Heuristic 1 describes the process of transitioning from the set of SINR intervals to the regression line represented by (10). The SINR per RNTI, in the absence of interference, is given by (11).

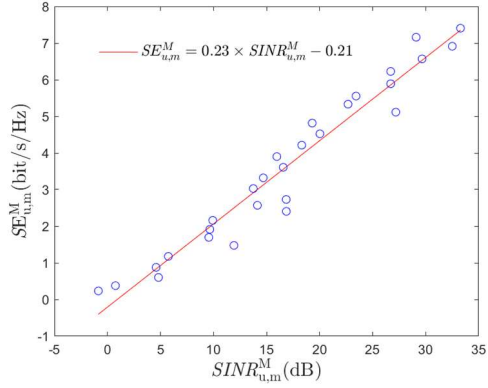

**Fig. 1.** Linear regression of the interpolated points for $SE_{u,m}^M$ versus $SINR_{u,m}^M$ as per (10).

$$SINR_{u,m}^M = 10\, log_{10}(\frac{R_{u,m}^M}{N_{fr1}}) \qquad (11)$$

The received power and thermal noise power for each RNTI are given by (12) and (13), respectively.

$$R_{u,m}^M = \frac{10^{(\frac{Pt-30}{10})}}{U_m\, PL_{UMa\_LOSu,m}^M} \qquad (12)$$

$$N_{fr1} = 10^{[-1\ .9+log_{10}(B\,\Delta fr1)]} \qquad (13)$$

The RSRP defined in (12) is based on 3GPP path loss model for an UMa scenario, whose cells are positioned at a height of 25 m, according to [23]. This configuration simplifies the problem to a 2D optimisation, considering the plane $z_m^M = h_{BS} = 25$ m, for all the $M$ cells.

The pathloss is modeled as a two-branch function by (14) with a non-linearity point at a 2D breakpoint distance between the UEs and cells.

$$PL^M_{UMa\_LOSu,m} = \begin{cases} PL^M_{1\,u,m}, & 10\,m \leq d^M_{2D,m} \leq d^M_{BPu,m} \\ PL^M_{2\,u,m}, & d^M_{BPu,m} \leq d^M_{2Du,m} \leq 5\,km \end{cases} \quad (14)$$

In UMa scenarios, this 2D breakpoint distance depends on the antenna heights, the probability of LoS, the effective environment height ($h_E$) and the carrier frequency. hE is set to 1 m, with a LoS probability of 1 / (1+C ($d^M_{2Du,m}$, $h_{UT}$)), which is 1 for user heights ($h_{UT}$) up to 13 m. According to the definition in [23], $PL^M_{UMa\_LOSu,m}$ is represented by $PL^M_{1\,u,m}$ for 2D distances below the breakpoint distance and takes the values of $PL^M_{2\,u,m}$ for 2D distances above that point, as presented in (15) and (16).

---

**Heuristic 1** – SE vs SINR linear regression leading to (10). *(Comments in gray refer to the lines below)*

1: */* Access the structured information containing the 28 SINR intervals, for each*
2: */* MCS, in [22] and [24]*
3: SINR_min(0 to 27) = **read** structure of min SINR values
4: SINR_max(0 to 27) = **read** structure of max SINR values
5: */* Convert the SINR values from dB to linear scale*
6: **for** MCS = 0 **to** 27
7:    SINR_min (MCS) = 10^(SINR_min (MCS)/10)
8:    SINR_max (MCS) = 10^(SINR_max (MCS)/10)
9:    */* For each SINR interval, calculate the average SINR and convert it back to*
10:   */* dB*
11:   SINR_avg (MCS) = 10log10[(SINR_max (MCS) – SINR_min (MCS))/2 +
12:   SINR_min (MCS)]
13:   */* Each MCS corresponds to a specific spectral efficiency value by 3GPP*
14:   */* definition, as in [22]*
15:   **get** spect_eff (MCS) value **from** [22]
16: */* Calculate the slope and intercept of the regression line that best fits the 28*
17: */* (SINR; Spectral Efficiency) points in Fig. 1*
18: **compute** slope, intercept **from** SINR_avg **and** spect_eff

---

$$PL^M_{1\,u,m} = 10^{2.8} d^{M\,2.2}_{3Du,m} f^{M\,2}_m \quad (15)$$

$$PL^M_{2\,u,m} = 10^{2.8} \frac{d^{M\,4}_{3Du,m} f^{M\,2}_m}{(d^{M\,2}_{BPu,m} + (hBS - hU)^2)^{0.9}} \quad (16)$$

The model excludes pathloss values for 2D distances above 5000 m, as these are beyond the range considered for *V*, and below 10 m, which is ensured by the optimisation constraint (3). The distances $d^M_{3Du,m}$, $d^M_{2Du,m}$, and $d^M_{BPu,m}$ are given by (17), (18) and (19), respectively.

$$d^M_{3Du,m} = \sqrt{(x^M_m - x^U_u)^2 + (y^M_m - y^U_u)^2 + (z^M_m - z^U_u)^2} \quad (17)$$

$$d^M_{2Du,m} = \sqrt{(x^M_m - x^U_u)^2 + (y^M_m - y^U_u)^2} \quad (18)$$

$$d_{BPu,m}^{M} = \frac{4\,(h_{BS} - h_E)(h_{UT} - h_E)\,f_m^M\,10^9}{c} \quad (19)$$

In 5G NR, the SCS for FR1, as referred to in (7), and the typical numerology $\mu_{fr1} = 0$, result in a bandwidth equal to 15 kHz per RB. With 9 sub-carriers per RB used for data (the remaining 3 are used for control) and 266 RBs per slot, the total bandwidth available for data transmission is equal to 266 x 9 x 15 kHz = 35.9 MHz.

The user association binary variable $S_{u,m}^M$, as defined in constraint (2), ensures that each $UE_u$ is associated with a single NGMC. The constraint in (3) establishes the lower bound of 10 m for the 2D distance between each $UE_u$ and the NGMC, ensuring the applicability of the 3GPP pathloss PL1 range, as described in [23]. Constraint (4) sets the upper limit for the spectral efficiency according to the definition in [24], considering table 2 of EESM 5G NR error model type, with chase combining. In the frequency domain, we have considered 9 sub-carriers per RB instead of 12; in the time domain, we have also considered 12 OFDM symbols per slot rather than 14, to account for overhead. Therefore, the spectral efficiency's upper bound is 6.4000 bit/s/Hz compared to the typical 7.4063 bit/s/Hz for an MCS index of 27 and modulation order of 8, as stated in [22].

### 3.2   Solving the problem

The optimisation problem requires solving the 3D Euclidean coordinates of the NGMCs $(x_m^M, y_m^M, z_m^M)$ and the set of Boolean user associations $S_{u,m}^M$ for each pair $UE_u$ and $NGMC_m$. In order to solve this optimisation problem, we have initialised the optimisation variables and used a state-of-the-art solver to iteratively converge to a solution. The Genetic Algorithm (GA), implemented in MATLAB, was employed to determine solutions for the various scenarios considered. GA is an optimisation solver inspired by the principles of natural selection and genetics [18]. It is widely used to solve complex optimisation problems where traditional mathematical approaches may be inefficient or impractical. However, the problem formulation is independent of the solver used; exploring different types of solvers is beyond the scope of this work and is recommended for future research.

## 4   Performance Evaluation

In this section, we introduce a network performance evaluation based on a series of process steps, detailed in Fig. 3, for each of the scenarios. The study focuses on Terminal XXI of Sines seaport, a future smart port with high wireless connectivity. We consider an environment with different sets of UEs, their centroid points or Geo-Means, and the maximum pairwise distance or diameter between them. For example, the NGMC's obtained position that best serves the UEs is shown in Fig. 2.

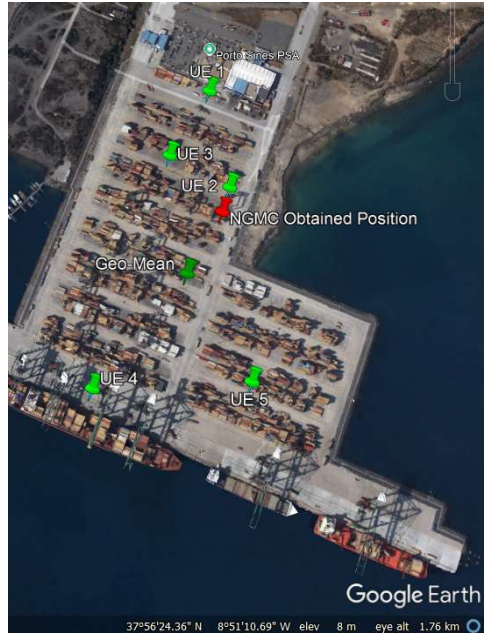

**Fig. 2.** Example of NGMC deployment at Terminal XXI of the Sines seaport, compared with the commonly used Geo-mean or Centroid position of the five UEs considered.

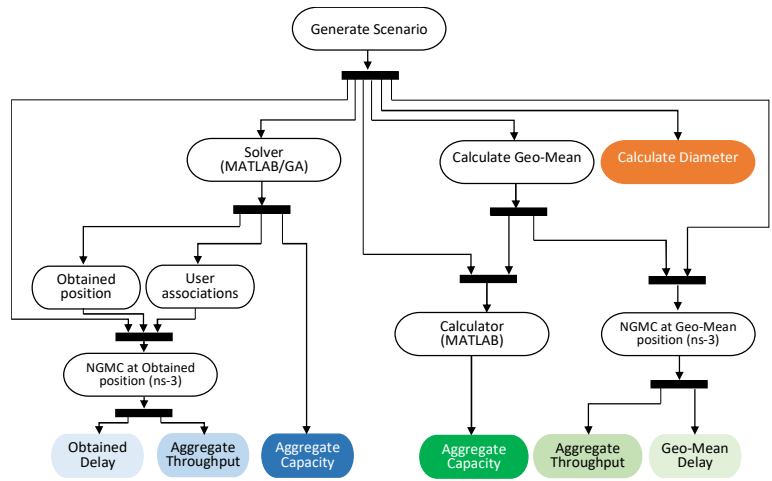

**Fig. 3.** Performance evaluation methodology for each scenario.

First, we obtained the NGMC position, user associations, and aggregate capacity using the optimisation problem in (1); this was developed with a maximum of 5000 generations and a tolerance of $10^{-30}$ for fast convergence and higher accuracy of the fitness function. Then, we simulated each scenario in ns-3, considering the parameters presented in Table 2, to obtain the aggregate throughput and delay. For each networking scenario, we calculated the mean geographical location of the UEs, their diameter and the respective aggregate capacity with the Calculator (MATLAB), as shown in Fig. 3.

This process was repeated in ns-3 to establish a baseline to compare aggregate capacity with throughput. Along the performance evaluation, we used a set of metrics and indicators, including:

- Aggregate capacity – sum of the $U$ RNTI capacities provided by an NGMC to all associated $UE_u$ within $V$. This value is calculated by both the solver and the calculator;
- Aggregate throughput – sum of the average throughput measured by the ns-3 *flow monitor* module on each RNTI's downlink packet flow. This output is obtained from each ns-3 simulation;
- Maximum pairwise distance – also called diameter, it measures the distance between the two most distant UEs;
- Delay – average delay across all $U$ RNTI downlink packet flows, measured by *flow monitor*. Each RNTI's downlink packet flow is provided by the NGMC and is an ns-3 simulation output for each scenario;
- Obtained position – NGMC's Euclidean position determined by the solver to optimise the aggregate capacity;
- Geo-mean position – NGMC's Euclidean position based on the geographical average of all considered $UE_u$ in a scenario. This was used as input to the calculator to obtain the baseline aggregate capacity.

In order to compare the results from the solver and the calculator with those from the ns-3 network simulator, we analysed a first scenario with two UEs at 1 km 2D distance from each other, as depicted in Fig. 4.

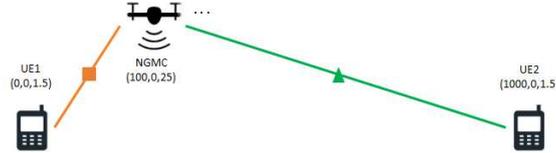

**Fig. 4.** First scenario with 2 UEs and an NGMC.

The solver output is confirmed at the extreme points of the scenario - (100, 0, 25) and (900, 0, 25) - as obtained positions with the highest aggregate capacity values. This is shown in Fig. 5, with step point coordinates depicted at the x-axis and respective results on y-axis. For each point, between a 100 m distance from UE1 to a 100 m distance from UE2, the aggregate capacity and aggregate throughput were obtained using the calculator and then validated in ns-3, with evaluations conducted at every 100 m step. The results show significant similarities between capacities and throughputs. The lines with round markers in Fig. 5 represent the sum of the flows below them. The upper blue continuous line represents the aggregate throughput of the two corresponding flows below, with 95% confidence intervals. Analysing the trajectory lines reveals that the aggregate throughput and capacity values at the geo-mean position (500, 0, 25) are lower compared to the obtained positions at the extremes. This is because the flows are not linear; they follow an exponential pattern.

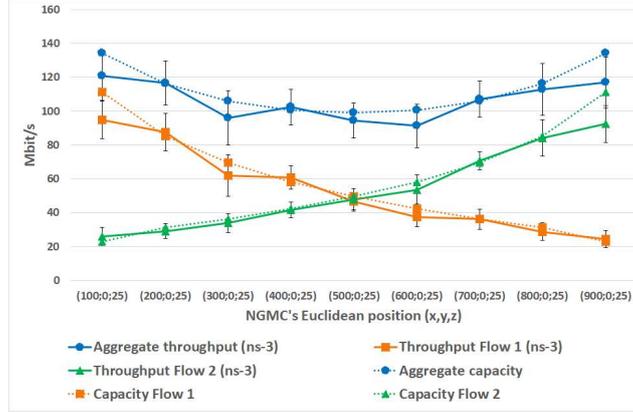

**Fig. 5.** Aggregate capacity, aggregate throughput and average results for each flow.

**Table 2.** Parameters used in the solver and ns-3 simulations.

| Parameter | Value |
| --- | --- |
| 3GPP Scenario | UMa |
| Number of NGMCs ($M$) | 1 |
| Number of UEs ($N$) | From 2 to 5 |
| Carrier frequency (FR1) | 5 GHz |
| Numerology ($\mu_{fr1}$) | 0 |
| SCS ($\Delta f_{r1}$) | 15 kHz |
| gNB and UE antenna model | Isotropic (SISO) |
| Transmission power | 24 dBm |
| Required data rate | 100 Mbit/s |
| Traffic direction | Downlink |
| Error Model | AMC with NrEesmCcT2 |
| NR MAC Scheduler Type | OFDMA Round Robin |
| Number of RBs | 266 |
| Number of Subcarriers per RB | 9 |

This paper studies several scenarios and uses the set of parameters presented in Table 2, which were applied to the solver, the calculator and the ns-3 simulator. As a preliminary study, this paper does not consider obstacle shadowing or any backhaul capacity limits. It uses an FR1 macro scenario with the simplest numerology and corresponding SCS, taking advantage of the maximum number of RBs in frequency per slot (266); however, some subcarriers are not considered for data (3), as they are assumed to be used for control or reference signals. Additionally, AMC technique in EESM error model was chosen to be simpler for HARQ processes with Chase Combining (NrEesmCcT2), rather than the more complex Incremental Redundancy (NrEesmIrT2) option. A detailed study of the NGMC for every point along the trajectory between the two UEs was made only for the first scenario. Since the results in Fig. 5 show significant similarities, we continued with additional scenarios, following these calibration tests, which validate our initial findings.

Fig. 6 presents results starting from first scenario, labeled as "Set 0". The other scenarios on the x-axis of Fig. 6 were studied with different sets of UEs, randomly generated and distributed across the plane at z = $h_{BS}$ of V. The maximum pairwise distances for each scenario are plotted on the rightmost y-axis showing an increase from Set 1 to Set 5. For each scenario, we followed the threads of process described in Fig. 3, leading to the metrics presented in the same colours in Fig. 6. Each ns-3 simulation was repeated ten times, each one considering 1000 subframes, 12000 OFDM symbols and 266000 RBs for numerology 0, allowing for significant variability. Results are presented in Fig. 6 with 95% confidence intervals.

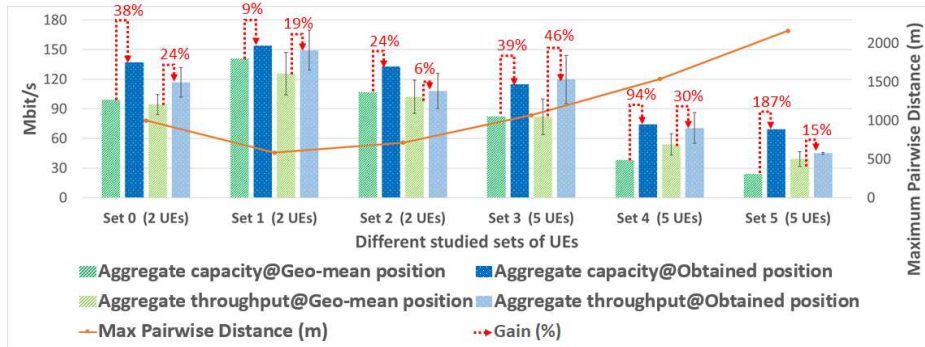

**Fig. 6.** Aggregate capacities and respective throughputs from ns-3 for obtained positions and geo-mean positions.

When considering the dark and light blue bars (second and fourth in each set) for the obtained position of the NGMC, we observe aggregate capacity values similar to the respective aggregate throughputs. Conversely, comparing the dark and light green bars (first and third in each set) for the geo-mean position of the NGMC, they are generally below the first two, although similar for each set. The blue bars consistently exceed the green, indicating that placing the NGMC at the obtained positions with the solver, allows for better results than at geo-mean positions. It is also worth mentioning that the bar heights decrease as the diameter increases, which means lower throughputs and capacities for UE sets with greater maximum pairwise distances.

Delay is not considered in the formulation of our problem. As such, optimising delay is outside the scope of this work; however, the ns-3 network simulator results for delay are presented in Fig. 7. These results, plotted alongside the scenarios' maximum pairwise distance, follow the process outlined in Fig. 3 that end up in delay measurements. Overall, delays increase with greater diameter distances, but delays corresponding to the obtained positions by the solver are consistently lower than those for the cells positioned at geo-mean locations.

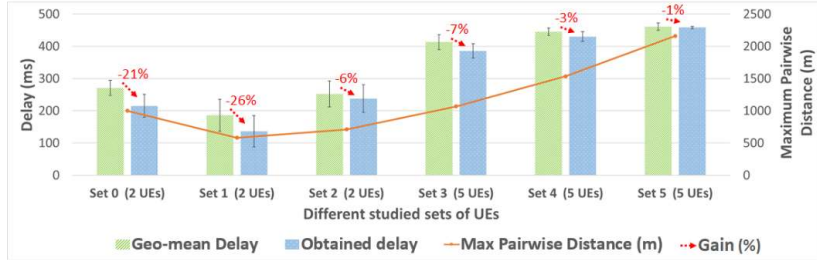

**Fig. 7.** Delays obtained on ns-3 for the obtained positions and geo-mean positions.

The gains in capacity, throughput and delay obtained by positioning the NGMC at obtained locations, when compared with geo-mean positions, are highlighted by red dotted arrows in Fig. 6 and Fig. 7. The gains reach up to 187% for capacity with obtained positioning (Set 5 in Fig. 6) when compared to geo-mean positioning: up to 46% for throughput (Set 3 in Fig. 6) and up to a 26% reduction for delay (Set 1 in Fig. 7). For scenarios with larger diameters, capacity gains are significantly higher, despite lower throughput and delay gains due to increased HARQ retransmissions.

Fig. 8 depicts the Cumulative Distribution Function (CDF) of the delays obtained from ns-3 simulations, which were conducted 10 times for each position. For each set of UEs studied, there are two CDFs: one for the obtained position of the NGMC (continuous lines) and the other for the geo-mean position (dotted lines). Scenarios with larger diameters present steeper delay distributions, while scenarios with smaller diameters present more dispersed distributions, indicating greater variability in delays. For the 90$^{th}$ percentile, delay reductions reach up to 30%, as depicted in Fig. 8 for Set 1.

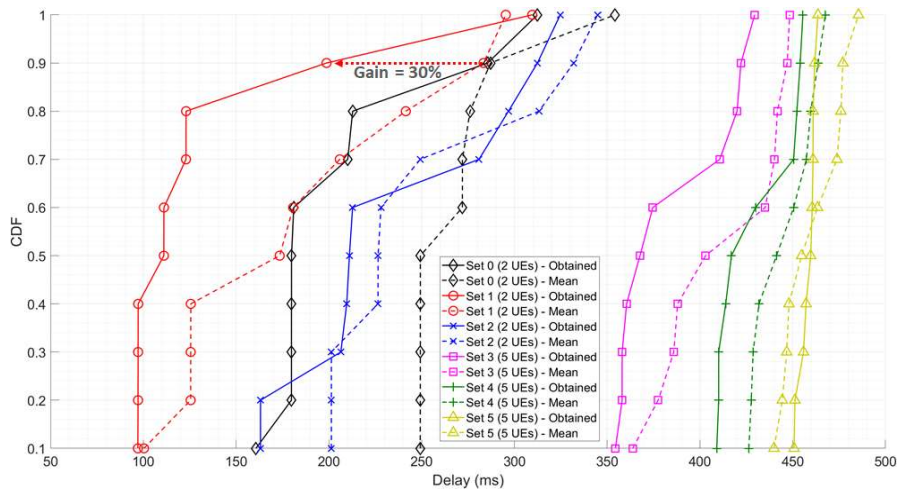

**Fig. 8.** CDF of the delays from the simulations.

## 5  Conclusions

In this paper, we propose a system model formulated as an optimisation problem with a heuristic to obtain the position for an NGMC node and a network performance evaluation study to compare results. The obtained position maximises the aggregate capacity for a set of UEs. The proposed system model and solver consistently determine obtained positions for the NGMC, achieving significant gains, including delay reductions. With this paper's contributions, we pave the way for improving dynamic radio network planning tasks for MNOs in next-generation outdoor environments, such as seaports with multiple UEs. This approach allows MNOs to optimise their investments in radio planning by focusing on performance rather than merely covering areas with geometrically distributed fixed cells.

As future work, we aim at addressing scenarios with a larger number of UEs and consider the blockage effect caused by obstacles, such as containers in seaports, which can limit signal propagation and reduce network capacity and throughput. Moreover, we aim at considering multiple mobile and fixed backhaul cells and explore different types of scheduling over available RBs with appropriate numerologies.

### Acknowledgements

This work is co-financed by Component 5 - Capitalization and Business Innovation, integrated in the Resilience Dimension of the Recovery and Resilience Plan within the scope of the Recovery and Resilience Mechanism (MRR) of the European Union (EU), framed in the Next Generation EU, for the period 2021 - 2026, within project NEXUS, with reference 53.

15University, Raleigh, NC, Department of Computer Science, University of Chicago, Chicago, IL, 2018

7. A. Fouda et al., "Interference Management in UAV-Assisted Integrated Access and Backhaul Cellular Networks", Department of Electrical and Computer Engineering, Florida International University, Miami, FL, Department of Electrical and Computer Engineering, North Carolina State University, Raleigh, NC, Department of Computer Science, University of Chicago, Chicago, IL, IEEE Access, 2019
8. C. Qiu et al., "Joint Resource Allocation, Placement and User Association of Multiple UAV-Mounted Base Stations with In-Band Wireless Backhaul", Key Laboratory of Universal Wireless Communications, Ministry of Education, Beijing University of Posts and Telecommunications, Beijing 100876, China, 2019
9. E. Kalantari et al., "User association and bandwidth allocation for terrestrial and aerial base stations with backhaul considerations", 2017 IEEE 28th Annual International Symposium on Personal, Indoor, and Mobile Radio Communications (PIMRC), 2017
10. A. Kubota et al., "A Study on Conversion of NLoS to LoS conditions using Sidelink in Smart Factory Environments", 2021 IEEE VTS 17th Asia Pacific Wireless Communications Symposium (APWCS), Osaka, Japan, 2021
11. V. Monteiro et al., "Paving the Way Toward Mobile IAB: Problems, Solutions and Challenges", IEEE Open Journal of the Communications Society (Volume: 3), 2022
12. J. Chen et al., "3D Urban UAV Relay Placement: Linear Complexity Algorithm and Analysis", IEEE Transactions on Wireless Communications, Vol. 20, No. 8, August 2021
13. J. Chen et al., "Efficient Local Map Search Algorithms for the Placement of Flying Relays", IEEE Transactions on Wireless Communications, Vol. 19, No. 2, February 2020
14. C.-H. Lee et al, "Mobile Small Cells for Further Enhanced 5G Heterogeneous Networks", ETRI Journal, Volume 37, Number 5, October 2015
15. A. Coelho et al., "Traffic-aware Gateway Placement for High-capacity Flying Networks", 2021 IEEE 93rd Vehicular Technology Conference, VTC2021-Spring
16. E. Almeida et al., "Joint traffic-aware UAV placement and predictive routing for aerial networks", Elsevier Ad Hoc Networks 118 (2021) 102525, INESC TEC and Faculdade de Engenharia, Universidade do Porto, Portugal, 2021
17. M. Shehzad et al., "Backhaul-Aware Intelligent Positioning of UAVs and Association of Terrestrial Base Stations for Fronthaul Connectivity", IEEE Transactions on Network Science and Engineering, Vol. 8, No. 4, October-December 2021
18. X. Zhong et al., "QoS-Compliant 3-D Deployment Optimization Strategy for UAV Base Stations", IEEE Systems Journal ( Volume: 15, Issue: 2, June 2021)
19. Y. Huang et al., "Joint AMC and Resource Allocation for Mobile Wireless Networks Based on Distributed MARL", 2022 IEEE International Conference on Communications Workshops (ICC Workshops), Seul, South Korea, 2022
20. F. Karo et al., "5G New Radio (NR) Network Planning at Frequency of 2.6 GHz in Golden Triangle of Jakarta", 3rd International Seminar on Research of Information Technology and Intelligent Systems (ISRITI), 2020
21. V. Agelliza et al., "Analysis of the implementing inter band carrier aggregation (ca) on the 5G new radio (NR) networks", Journal of information technology and its utilization, volume 6, issue 1, Indonesia, June-2023
22. 3GPP TS 38.214 V16.5.0, Table 5.1.3.1-2
23. 3GPP TR 38.901 version 17.0.0 Rel. 17, Table 7.4.1-1
24. S. Lagen et al., "New Radio Physical Layer Abstraction for System-Level Simulations of 5G Networks", CTTC, Interdigital